\documentclass[runningheads]{llncs}
\usepackage{caption}
\usepackage{xcolor,chngcntr}
\usepackage{multirow}

\AtEndEnvironment{listing}{\vspace{-24pt}}

\usepackage{gensymb}
\usepackage{amsfonts}
\usepackage{hyperref}
\usepackage[T1]{fontenc}
\usepackage{graphicx}
\usepackage{color}
\begin{document}
\title{Spectral U-Net: Enhancing Medical Image Segmentation via
Spectral Decomposition}
\titlerunning{Spectral U-Net}
%
\author{Yaopeng peng, Milan Sonka, Danny Z. Chen}
\authorrunning{Anonymous}
%
\institute{}
\maketitle              
\begin{abstract}
This paper introduces Spectral U-Net, a novel deep learning network based on spectral decomposition, by exploiting Dual Tree Complex Wavelet Transform (DTCWT) for down-sampling and inverse Dual Tree Complex Wavelet Transform (iDTCWT) for up-sampling. We devise the corresponding Wave-Block and iWave-Block, integrated into the U-Net architecture, aiming at mitigating information loss during down-sampling and enhancing detail reconstruction during up-sampling. In the encoder, we first decompose the feature map into high and low-frequency components using DTCWT, enabling down-sampling while mitigating information loss. In the decoder, we utilize iDTCWT to reconstruct higher-resolution feature maps from down-sampled features. Evaluations on the Retina Fluid, Brain Tumor, and Liver Tumor segmentation datasets with the nnU-Net framework demonstrate the superiority of the proposed Spectral U-Net. 


\end{abstract}

\section{Introduction}
Deep neural networks (DNNs) have emerged as a predominant approach for medical image segmentation, which is a task assigning a class label to each image pixel. U-Net~\cite{ronneberger2015u} is a Convolutional Neural Network (CNN) that extracts multi-scale features of varying resolutions in the encoding stage and progressively reconstructs the resolutions in the decoding stage. In the encoding stage, down-sampling serves to sub-sample the feature maps generated by the convolutional layers, hence reducing the spatial resolutions of the feature maps. This reduction aims to alleviate the computational burden and enhance the overall efficiency of the network. Down-sampling aggregates information from neighboring pixels, capturing features of varying granularity and scales. Moreover, down-sampling assists the network to mitigate its sensitivity to minor variations in feature positions of the image and to small noise in the features, thus fostering robustness of the extracted feature maps. Besides, down-sampling enlarges the receptive field of neurons in deeper layers, enabling neurons to glean information from a broader region of the input image, consequently empowering the network to discern more intricate patterns and relationships. During the decoding stage, up-sampling is utilized to restore the original spatial resolution, thereby enabling the network to reconstruct detailed information for pixel level predictions.

As a fundamental process of DNNs, multiple down-sampling mechanisms have been proposed, among which the most prevalent ones include max pooling~\cite{yamaguchi1990neural}, average pooling~\cite{lecun1989handwritten}, mixed pooling~\cite{yu2014mixed}, stochastic pooling~\cite{zeiler2013stochastic}, and strided convolution~\cite{lecun1989handwritten}. Despite the simplicity and effectiveness, these down-sampling techniques may induce discontinuities and loss of details, especially when dealing with small objects such as tumors in medical images, thus impeding accurate reconstruction of spatial resolutions during the up-sampling process.

To address the aforementioned challenge, we introduce a new framework called Spectral U-Net, aiming to extract hierarchical feature maps of varying resolutions for medical image segmentation. Our framework seeks to alleviate information loss during the down-sampling and up-sampling processes. Our main contributions are two-fold.  (1) We build a Wave-Block in the encoding stage for performing down-sampling. We first decompose the input feature map into two components, with low and high frequencies respectively, utilizing the Dual Tree Complex Wavelet Transform (DTCWT)~\cite{selesnick2005dual}. These components are then processed with shared convolutional fileters, resulting in a reduction in spatial resolution and an increase in channel. This approach enables the network to retain the benefits of pooling while mitigating the issue of information loss. (2) We devise an inverse wavelet block (iWave-Block) in the decoding stage for resolution reconstruction. The iWave-Block employs the inverse Dual Tree Complex Wavelet Transform (iDTCWT) to reconstruct the original input from the down-sampled wavelet coefficients, retaining all pertinent information while reconstructing spatial resolutions, thus mitigating information loss. 

We evaluate our new method on several datasets (e.g., Retina Fluid~\cite{bogunovic2019retouch}, BRATS 2017~\cite{menze2014multimodal}, and LiTS 2017~\cite{bilic2023liver}) based on the nnU-Net framework~\cite{isensee2021nnu}, demonstrating the superiority of our approach compared to previous methods.

\vspace{-0.11in}
\section{Related Work}

\vspace{-0.05in}
\subsection{Down-sampling and Up-sampling}

To reduce the spatial resolutions of feature maps for computational efficiency and extract multi-scale features, down-sampling is utilized. Max pooling~\cite{yamaguchi1990neural} and average pooling~\cite{lecun1989handwritten} are two commonly-used methods that select the largest and mean values over a pooling region, respectively. Max pooling discards weaker signals within the pooling region, potentially ignoring some local details. Meanwhile, average pooling calculates the average of the pooling region, which can result in down-weighting strong activations. In~\cite{yu2014mixed}, a mixed pooling operation was proposed to combine max pooling and average pooling to improve image classification accuracy. In~\cite{zeiler2013stochastic}, a stochastic pooling scheme was proposed to randomly select pixels in the pooling region according to a multinomial distribution, which is determined by activities within the pooling region. Strided convolution~\cite{lecun1989handwritten} is another parameterized method for down-sampling feature maps.

\vspace{-0.1in}
\subsection{Wavelet Transform in Deep Learning}

Wavelet transform has been extensively employed in processing time-series signals and in pre-processing and post-processing of images. The Discrete Wavelet Transform (DWT) decomposes input signals into multiple coefficients possessing distinct frequencies and scales. Subsequently, each such coefficient is processed independently and recombined using the inverse Discrete Wavelet Transform (iDWT) to reconstruct the original signal or feature. In~\cite{bae2017beyond}, it proposed utilizing wavelet sub-bands to enhance the performance of image restoration.  A method was presented to recover high-resolution image details by leveraging sub-bands from low-resolution input~\cite{guo2017deep}. In~\cite{liu2018multi}, it gave a multi-level wavelet transform for expanding the receptive field for image restoration tasks.  Wavelet pooling was proposed to conduct a second-level decomposition of features, discarding the first-level sub-bands to reduce feature dimensions in image classification~\cite{williams2018wavelet}. In~\cite{fujieda2018wavelet}, wavelet transform was used to leverage spectral information for texture classification and image annotation. In~\cite{li2021wavecnet}, CNNs were integrated with wavelets for noise-robust image classification. In~\cite{yao2022wave}, wavelet transform was utilized to decrease the spatial resolution of key-value pairs in vision Transformer, thus reducing computational costs. However, the capacity of these methods to decompose low-frequency and high-frequency components of an image is still constrained, and their capability to capture fine-grained image details is inadequate.

In this paper, we incorporate the Dual Tree Complex Wavelet Transform (DTCWT)~\cite{selesnick2005dual} into a CNN framework to perform down-sampling and up-sampling, aiming to alleviate potential information loss associated with traditional methods such as max pooling and strided convolution. Specifically, we construct a Wave-Block and perform DTCWT along each direction of the feature map, thereby achieving lossless spatial resolution reduction. For the reconstruction stage, we introduce an iWave-Block and utilize the inverse DTCWT (iDTCWT) to reconstruct features based on the output of the previous level.

\begin{figure}[t]
\centering
\includegraphics[width=0.9\columnwidth]{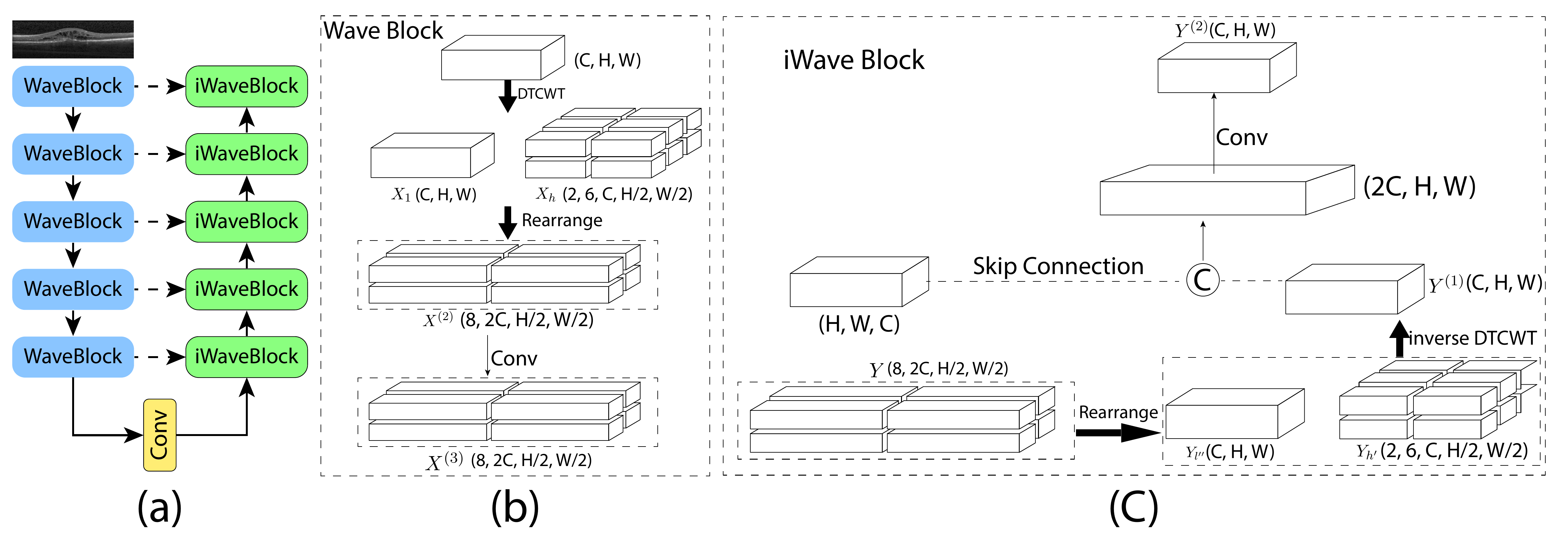}
\caption{(a) The U-shape structure of our proposed Spectral U-Net, which applies DTCWT (b) in the encoding stage for spatial resolution reduction and iDTCWT (c) in the decoding stage for resolution reconstruction.}
\label{fig:spec_net}
\end{figure}

\vspace{-0.15in}
\section{Method}

In this section, we first introduce the Dual Tree Complex Wavelet Transform (DTCWT)~\cite{selesnick2005dual} and our proposed Wave-Block and iWave-Block. The overall structure of our proposed framework is shown in Fig.~\ref{fig:spec_net}(a). In the encoding stage, we use the Wave-Block to incrementally down-sample the feature map while increasing the channels. In the decoding stage, we use the iWave-Block to progressively reconstruct the spatial resolution for subsequent segmentation.

\vspace{-0.15in}
\subsection{Dual Tree Complex Wavelet Transform}

The Discrete Wavelet Transform (DWT) 
decomposes a given signal into a number of sets, each set representing a sub-band of coefficients describing the signal in the corresponding frequency band. DWT often suffers from shift invariance and lacks directionality in two and higher dimensions. Dual Tree Complex Wavelet Transform (DTCWT), an extension of DWT, offers advantages over DWT regarding shift invariance and directional selectivity. It first decomposes an input feature $X$ into a low-frequency and a high-frequency components $F_l$ and $F_h$:
\begin{equation}
X \rightarrow F_l + F_h = c\, \phi + \sum_{m=1}^M\sum_{k=1}^6 d_m^k \Phi,
\end{equation}
where $\phi$ and $\Phi$ denote the wavelet transforms for the low and high frequencies respectively, $c$ and $d_m^k$ denote the corresponding filters, $M$ is the number of levels of the decomposition, and $k$ refers to directional selectivity.

\vspace{-0.15in}
\subsection{Wave-Block}
\label{wave_sec}

In DNNs, max pooling and average pooling are both irreversible and can cause information loss. To mitigate this issue, we design an invertible down-sampling block, $WaveBlock$, which performs invertible down-sampling by utilizing the spectral decomposition, i.e., DTCWT, as shown in Fig.~\ref{fig:spec_net}(b).

For an input feature map $X\in \mathbb{R}^{C\times H\times W}$, where $C$, $H$, and $W$ denote the channels, height, and width of $X$ respectively, DTCWT first decomposes $X$ into a low-frequency component $X_l$ and a high-frequency component $X_h$, with $X_l\in \mathbb{R}^{C\times H\times W}$ and $X_h\in \mathbb{R}^{2\times 6 \times C\times \frac{H}{2}\times \frac{W}{2}}$, where $2$ denotes the real and imaginary parts, and $6$ represents the six orientations of DTCWT spectral transform: $15\degree$, $45\degree$, $75\degree$, $105\degree$, $135\degree$, and $165\degree$. 
Then, we rearrange $X_l$ into $X_{l'}$ using the pixel shuffle mechanism~\cite{shi2016real}, with $X_{l'}\in \mathbb{R}^{4\times C\times \frac{H}{2}\times \frac{W}{2}}$. Next, we rearrange $X_{l'}$ and $X_h$:
\begin{equation}
X^{(2)} = rearrange(X_{l'}, X_{h}),
\end{equation}
where $X^{(2)} \in \mathbb{R}^{8\times 2C\times \frac{H}{2}\times \frac{W}{2}}$, which is then processed with a convolutional block:
\begin{equation}
X^{(3)} =  \sigma(BN(X^{(2)}\cdot w_1)),
\end{equation}
\begin{figure}[h!]

\centering
\includegraphics[width=1\linewidth]{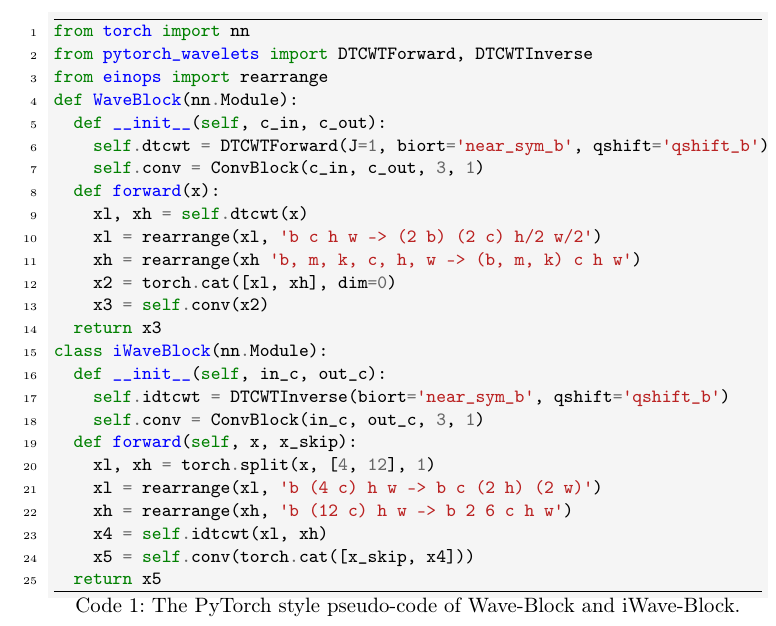}
\caption*{}
\label{code1}
\end{figure}
where $X^{(3)}\in \mathbb{R}^{8\times 2C\times \frac{H}{2}\times \frac{W}{2}}$, $w_1$ denotes the parameters of the convolution, and $BN$ and $\sigma$ denote batch normalization and ReLU activation respectively. Through a Wave-Block, we have down-sampled the spatial resolution of $X$ and doubled the channels. The output $X^{(3)}$ is then sent to the next Wave-Block for further processing. PyTorch style pseudo-code of Wave-Block is shown in the following Code 1:


\vspace{-0.15in}
\subsection{iWave-Block}

In the decoding stage, we progressively reconstruct the down-sampled feature map back to the original spatial resolution, as shown in Fig.~\ref{fig:spec_net}(c).

Given the down-sampled feature map $Y\in \mathbb{R}^{8\times 2C \times \frac{H}{2}\times \frac{W}{2}}$, we first rearrange it into the low and high frequency components $Y_{l'}$ and $Y_{h'}$, with $Y_{l'}\in \mathbb{R}^{4\times C\times \frac{H}{2} \times \frac{W}{2}}$ and $Y_{h'} \in \mathbb{R}^{2\times 6\times C\times \frac{H}{2}\times \frac{W}{2}}$. $Y_{l'}$ is then rearranged into $Y_{l''}$:
\begin{equation}
Y_{l''} = rearrange(Y_{l'}),
\end{equation}
with $Y_{l''}\in \mathbb{R}^{C\times H\times W}$. Afterwards, we reconstruct the spatial resolution with the inverse DTCWT transform:
\begin{equation}
Y^{(1)} = idtcwt(Y_{l''}, Y_{h'}),
\end{equation}
where $idtcwt$ denotes the inverse DTCWT transform, and $Y^{(1)}\in \mathbb{R}^{C\times H\times W}$. Next, we concatenate $Y^{(1)}$ and $X_{skip}$ (the feature map from the skip connection) along the channel dimension, and feed the concatenation to a convolution layer:
\begin{equation}
Y^{(2)} = \sigma(BN(Concat(Y^{(1)}, X_{skip})\cdot w_2)),
\end{equation}
where $\sigma$ denotes ReLU activation, $BN$ denotes batch normalization, $w_2$ denotes the parameters of the convolution, and $Y^{(2)}\in \mathbb{R}^{C\times H\times W}$. After the iWave-Block processing, we have up-sampled the spatial resolution of the map $Y$ and halved the channels. Hence, an iWave-Block can be considered as an up-sampling process. The PyTorch style pseudo-code of iWave-Block is given in Code~\ref{code1}.

\vspace{-0.15in}
\section{Experiments}

\vspace{-0.1in}
\subsection{Datasets}
We evaluate our proposed Spectral U-Net on the following datasets.

\noindent
\textbf{Retina Fluid}: We use the public RETOUCH and our in-house datasets. The label classes are intraretinal fluid (IRF), subretinal fluid (SRF), and pigment epithelial detachment (PED). The RETOUCH dataset is from three OCT scanners:  Cirrus, Spectralis, and Topcon, with 24, 24, and 22 subjects, respectively. Our in-house dataset contains 100 OCT subjects from the Spectralis scanner. The Retina Fluid dataset used is the union of the RETOUCH 
and in-house datasets.

\noindent
\textbf{BRATS 2017}: An MRI dataset for brain tumor segmentation. It comprises multimodal MRI scans: T1-weighted, T1-weighted contrast-enhanced, T2-weighted, and FLAIR images. The targets include three tumor sub-regions: enhancing tumor, peritumoral edema, and necrotic and non-enhancing tumor core. The annotations were combined into three nested sub-regions: Whole Tumor (WT), Tumor Core (TC), and Enhancing Tumor (ET).  

\noindent
\textbf{LiTS 2017}: This dataset uses 131 contrast-enhanced abdominal CT volumes 
for training/validation, and 70 for testing. 
Annotations include liver and tumor. 

\vspace{-0.1in}
\subsection{Experimental Setup}
All the experiments are conducted on an NVIDIA RTX 6000 Ada GPU. We test all the models based on the nnU-Net~\cite{isensee2021nnu} framework. Specifically, we optimize the models using the Gradient Descent optimizer with a momentum of 0.99. We employ a polynomial learning rate decay with a power of 0.9, and the initial learning rate is set to 0.01. We randomly split each dataset into training, validation, and test sets, with a ratio of 0.7:0.1:0.2. We report DSC (Dice Similarity Coefficient) and HD95 (Hausdorff Distance (95th percentile)) scores. Each experiment is run 5 times, and the averaged results are reported.

\vspace{-0.1in}
\subsection{Results}
Table~\ref{tab:retina} shows experimental results on the Retina Fluid dataset. We compare our method with several typical methods: (1) Curvature~\cite{xing2022multi}; (2) MsTGANet~\cite{wang2021mstganet}; (3) CPFNet~\cite{feng2020cpfnet}; (4) DconnNet~\cite{yang2023directional}; (5) Swin UNETR~\cite{hatamizadeh2021swin}; (6) nnU-Net~\cite{isensee2021nnu}. As shown, our Spectral U-Net improves the DSC scores for SRF and PED by 1.28\% and 2.35\% over nnU-Net respectively, and its DSC score for IRF is comparable to that of nnU-Net (0.08\% lower), demonstrating the effectiveness of our method in preserving information during down-sampling and up-sampling feature maps.









\setlength{\tabcolsep}{1pt}
\begin{table}[h!]
\scriptsize
\centering
\begin{tabular}
{c|c|c|c|c|c|c}
\multirow{2}{*}{Method} & \multicolumn{2}{c|}{IRF} & \multicolumn{2}{c|}{SRF} & \multicolumn{2}{c}{PED}   \\

\cline{2-7}
& DSC (\%)$\uparrow$ & HD95 (mm)$\downarrow$ & DSC (\%)$\uparrow$ & HD95 (mm)$\downarrow$ & DSC (\%)$\uparrow$ & HD 95 (mm)$\downarrow$ \\

\hline
Curvature~\cite{xing2022multi} & 79.23$\pm$1.50&0.089$\pm$0.001 & 82.89$\pm$1.54&0.084$\pm$0.001 & 83.04$\pm$1.82 & 0.663$\pm$0.003\\

MsTGANet~\cite{wang2021mstganet} & 78.61$\pm$1.98&0.091$\pm$0.001&83.01$\pm$1.59 & 0.089$\pm$0.001 & 82.26$\pm$1.71 & 0.678$\pm$0.004\\

CPFNet~\cite{feng2020cpfnet} & 79.77$\pm$1.59 & 0.078$\pm$0.001 & 83.79$\pm$1.64& 0.085$\pm$0.002 & 83.12$\pm$1.87 & 0.621$\pm$0.003\\

DconnNet~\cite{yang2023directional} & 81.28$\pm$1.59 & 0.068$\pm$0.001 & 84.51$\pm$1.54 & 0.078$\pm$0.001&84.12$\pm$1.47 & 0.612$\pm$0.004 \\

Swin UNETR~\cite{hatamizadeh2021swin} & 83.31$\pm$1.29 & 0.045$\pm$0.004 & 87.99$\pm$1.39 & 0.050$\pm$0.002 & 83.58$\pm$1.49 & 0.620$\pm$0.003\\

nnU-Net~\cite{isensee2021nnu} & \textbf{84.26$\pm$1.32} & \textbf{0.033$\pm$0.002} & 88.51$\pm$1.05&0.043$\pm$0.002 & 82.30$\pm$1.82&0.689$\pm$0.005\\

\hline
Spectral U-Net & 84.18$\pm$1.21 & 0.036$\pm$0.003 & \textbf{89.79$\pm$0.89} & \textbf{0.037$\pm$0.003} & \textbf{84.65$\pm$1.64} & \textbf{0.588$\pm$0.003}
\end{tabular}
\caption{Experimental results on the Retina Fluid dataset.
}
\label{tab:retina}
\end{table}

Table~\ref{ablation:brain} shows comparison results on the BRATS 2017 dataset with several typical methods: (1) TransBTS~\cite{wang2021transbts}; (2) SegResNet~\cite{myronenko20193d}; (3) VT-UNet~\cite{peiris2022robust}; (4) Swin UNETR~\cite{hatamizadeh2021swin}; (5) nnU-Net~\cite{isensee2021nnu}. We find that the DSC scores for ET and WT are improved  over Swin UNETR~\cite{hatamizadeh2021swin} by 0.32\% and 0.78\%, respectively. Compared to the improvements of Swin UNETR over nnU-Net (i.e., 0.11\% and -0.04\%), our improvements consistently demonstrate the capability of our Spectral U-Net to preserve information during the down-sampling and up-sampling processes.

Table~\ref{tab:liver} gives comparison results on the LiTS 2017 dataset with several typical methods: (1) TransBTS~\cite{wang2021transbts}; (2) KiU-Net~\cite{valanarasu2021kiu}; (3) Swin UNETR~\cite{hatamizadeh2021swin}; (4) nnU-Net~\cite{isensee2021nnu}. Our Spectral U-Net improves over the second-best method, Swin UNETR, the DSC scores for liver cancer and liver by 0.7\% and 0.45\%, respectively. Compared to the improvements of Swin UNETR over nnU-Net (i.e., 0.23\% and 0.34\%), our improvements underscore the capability of our method in mitigating information loss during the feature map down-sampling/up-sampling processes.

\setlength{\tabcolsep}{1pt}
\begin{table}[t]
\centering
\scriptsize
\begin{tabular}{c|c|c|c|c|c|c}
\multirow{2}{*}{Method} & \multicolumn{2}{c|}{ET} & \multicolumn{2}{c|}{TC} & \multicolumn{2}{c}{WT}  \\
\cline{2-7}
& DSC (\%)$\uparrow$ & HD95 (mm)$\downarrow$ & DSC (\%)$\uparrow$ & HD95 (mm)$\downarrow$ & DSC (\%)$\uparrow$ & HD95 (mm)$\downarrow$\\

 \hline
TransBTS~\cite{wang2021transbts} & 91.10$\pm$0.38& 7.379$\pm$0.025&93.87$\pm$0.21&4.915$\pm$0.035&90.89$\pm$0.30&6.913$\pm$0.041\\

SegResNet~\cite{myronenko20193d} & 92.12$\pm$0.40&6.398$\pm$0.022&94.68$\pm$0.24&3.960$\pm$0.023&91.20$\pm$0.25&6.545$\pm$0.039\\

VT-UNet~\cite{peiris2022robust} & 92.23$\pm$0.29 & 6.723$\pm$0.022 & 94.21$\pm$0.17 & 4.224$\pm$0.034 & 90.87$\pm$0.20&7.134$\pm$0.037\\

Swin UNETR~\cite{hatamizadeh2021swin} & 92.31$\pm$0.32 & 6.251$\pm$0.028 & \textbf{94.82$\pm$0.23} & \textbf{3.001$\pm$0.027} & 91.18$\pm$0.17&6.221$\pm$0.049\\
 
nnU-Net~\cite{isensee2021nnu} & 92.20$\pm$0.35 & 6.620$\pm$0.021&94.64$\pm$0.21&3.795$\pm$0.019&91.14$\pm$0.22&6.236$\pm$0.045\\ 
 
Spectral U-Net & \textbf{92.63$\pm$0.38} & \textbf{6.030$\pm$0.027} & 94.79$\pm$0.24 & 3.525$\pm$0.020 & \textbf{91.96$\pm$0.23}&\textbf{5.732$\pm$0.014} \\
\end{tabular}
\caption{Experimental results on the BRATS 2017 dataset.
}
\label{ablation:brain}
\end{table}

\setlength{\tabcolsep}{4pt}
\begin{table}[t!]
\centering
\scriptsize
\begin{tabular}{c|c|c|c|c}
\multirow{2}{*}{Method} & \multicolumn{2}{c|}{Liver} & \multicolumn{2}{c}{Cancer} \\
\cline{2-5}
& DSC (\%)$\uparrow$ & HD95 (mm)$\downarrow$ & DSC (\%)$\uparrow$ & HD95 (mm)$\downarrow$ \\

\hline
TransBTS~\cite{wang2021transbts} & 95.01$\pm$1.53&4.95$\pm$1.47&64.03$\pm$6.02&9.98$\pm$1.97\\

KiU-Net~\cite{valanarasu2021kiu} & 95.73$\pm$1.24&4.38$\pm$1.17&65.28$\pm$5.78&9.60$\pm$1.59\\

Swin UNETR~\cite{hatamizadeh2021swin} & 96.28$\pm$1.08&\textbf{3.71}$\pm$0.87&66.34$\pm$5.24&9.03$\pm$1.42\\

nnU-Net~\cite{isensee2021nnu} & 96.05$\pm$1.28 & 4.01$\pm$0.99  & 66.00$\pm$5.62&9.61$\pm$1.29\\ 

Spectral U-Net & \textbf{96.73$\pm$1.12} & 3.88$\pm$1.06& \textbf{67.04$\pm$5.25}&\textbf{8.77$\pm$1.38}\\

\end{tabular}
\caption{Experimental results on the LiTS 2017 dataset.
}
\label{tab:liver}
\end{table}

\vspace{-0.1in}

\subsection{Ablation Study}

We conduct ablation study to examine the effectiveness of our approach on the Retina Fluid dataset. The DSC scores for IRF, SRF, and PED are reported in Table~\ref{tab:ablation}. ConvBlock refers to traditional max-pooling convolutional block, and Linear-I denotes linear interpolation. We observe that our proposed Wave-Block consistently improves the DSC scores compared to ConvBlock. Note that the improvement when using only ConvBlock and iWave-Block together is limited since the information is already lost during the max-pooling process, making it unlikely to recover in the up-sampling process. This further demonstrates the effectiveness of using our DTCWT to mitigate information loss during down-sampling. Additionally, we compare DTCWT with the commonly-used Haar Wavelet transform~\cite{haar1909theorie}. The experimental results show that DTCWT performs better in decomposing high- and low-frequency components, thus better preserving information during the down-sampling process.

\setlength{\tabcolsep}{1pt}
\begin{table}[t!]
\centering
\scriptsize
\begin{tabular}{c|cccc|c|c|c}
Wavelets &  ConvBlock & Wave-Block & Linear-I  & iWave-Block & IRF (\%) & SRF (\%) & PED (\%)  \\

\hline
\multirow{4}{*}{Haar~\cite{haar1909theorie}} & \checkmark & & \checkmark  &  & \textbf{84.26$\pm$1.32} & 88.51$\pm$1.05 & 82.30$\pm$1.82 \\

&  & \checkmark & \checkmark  & & 84.13$\pm$1.17 &88.94$\pm$1.16 & 83.21$\pm$1.75 \\

& \checkmark &  &   & \checkmark & 84.21$\pm$1.20 & 88.48$\pm$0.87 & 82.35$\pm$1.60 \\

&  & \checkmark & & \checkmark & 83.96$\pm$1.19 & 89.05$\pm$0.95 &83.54$\pm$1.47 \\

\hline

\multirow{4}{*}{DTCWT}& \checkmark & & \checkmark  &  & \textbf{84.26$\pm$1.32} & 88.51$\pm$1.05 & 82.30$\pm$1.82  \\

&  & \checkmark & \checkmark & & 84.02$\pm$1.05&89.39$\pm$1.14 & 83.94$\pm$1.50 \\

& \checkmark &  &   & \checkmark & 84.13$\pm$1.23 & 88.49$\pm$1.21 & 82.85$\pm$1.45 \\

&  & \checkmark &  & \checkmark &  84.18$\pm$1.21 & \textbf{89.79$\pm$0.89} & \textbf{84.65$\pm$1.64} \\

\end{tabular}
\caption{Ablation study on the Retina Fluid dataset. ConvBlock refers to traditional max-pooling convolutional block, and
Linear-I denotes linear interpolation. }
\label{tab:ablation}
\end{table}

\vspace{-0.1in}

\subsection{Qualitative Results}
\begin{figure}[t!]
\centering
\includegraphics[width=0.8\columnwidth]{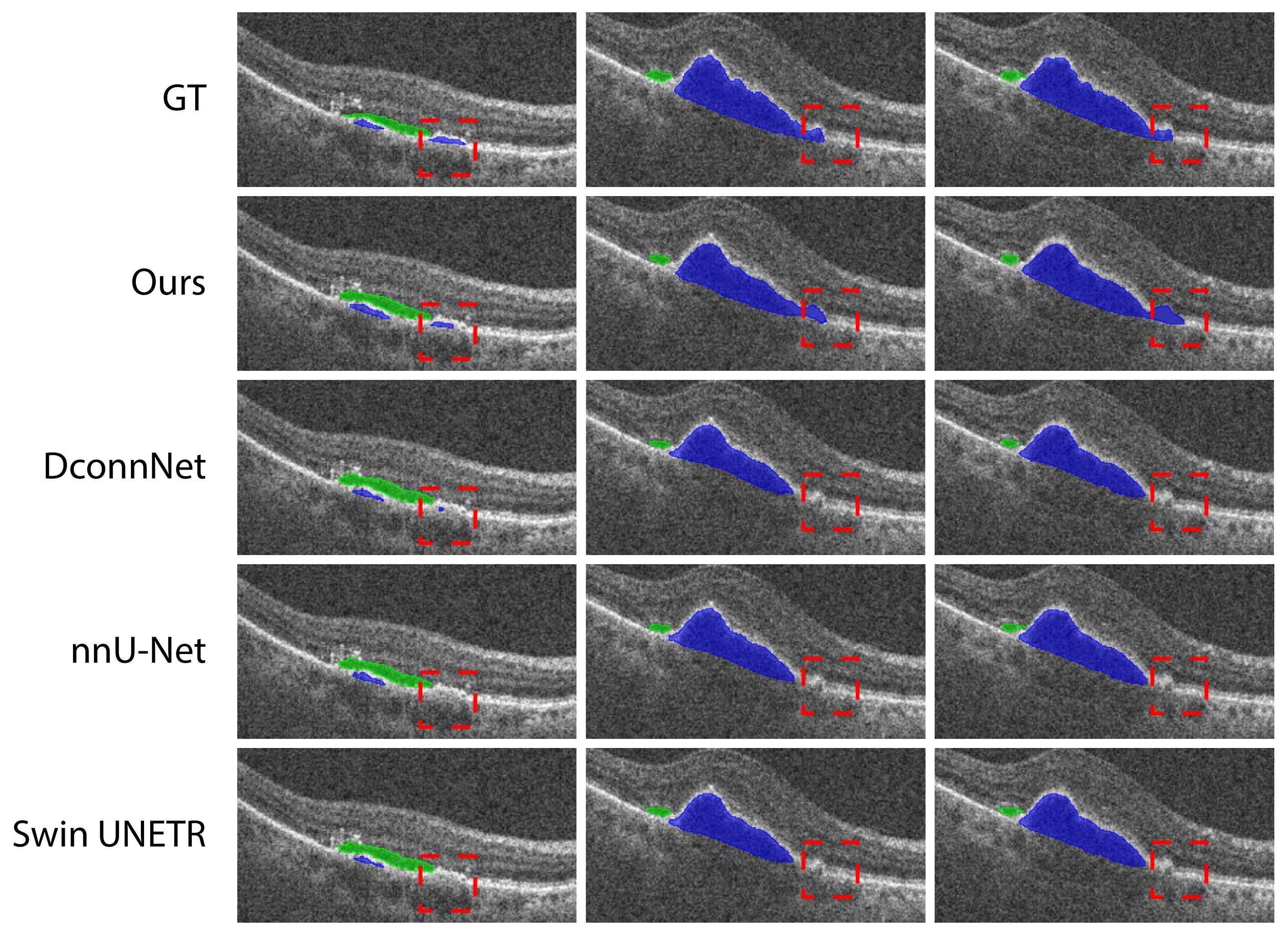}
\caption{Visual examples from the Retina Fluid dataset (\textcolor{green}{green} for SRF, \textcolor{blue}{blue} for PED). The dashed red boxes highlight that our method is able to capture intricate details of small objects and peripheral regions that are missed by nnUNet, Swin UNETR, and DconnNet.}
\label{fig:examples}
\end{figure}


Some qualitative examples from the Retina Fluid dataset are presented in Fig.~\ref{fig:examples}. These examples demonstrate that our method is capable of capturing intricate details of small objects and peripheral regions (marked by dashed red boxes), which are missed by nnUNet, Swin UNETR, and DconnNet.

\vspace{-0.1in}
\subsection{Parameters and Computation Costs}

To examine the costs of our method, we report the parameters, FLOPs (Floating Point operations), and Vol\_time (seconds to perform one volume prediction) in Table~\ref{tab:computation}. One can see that our parameters, FLOPs, and inference time are comparable to nnU-Net, showing the computational efficiency of our method.

\begin{table}[t!]
\scriptsize
\centering
\begin{tabular}{c|c|c|c}
Method & \# Params. & FLOPs & Vol\_time \\
\hline

nnU-Net & 18.225 M & 28.572 G & 10.24 s \\
Spectral U-Net & 18.593 M& 29.825 G & 10.52 s\\
\end{tabular}
\caption{Comparison of parameters and costs between Spectral U-Net and nnU-Net.}
    \label{tab:computation}
\end{table}

\vspace{-0.1in}
\section{Conclusions}

In this paper, we proposed a novel U-Net type segmentation network, Spectral U-Net, which utilizes Dual Tree Complex Wavelet Transform (DTCWT) for information lossless down-sampling and inverse Dual Tree Wavelet Transform (iDTCWT) for the up-sampling process. We used DTCWT to decompose feature maps into low- and high-frequency components in the encoding stage for down-sampling and used iDTCWT for reconstruction in the decoding stage, thus mitigating information loss of traditional max-pooling. Experimental results on the Retina Fluid, BRATS 2017, and LiTS 2017 segmentation datasets demonstrated the effectiveness of our new method while preserving computational efficiency.
%
%
%
\clearpage
\bibliographystyle{splncs04}
\bibliography{bibs}

\end{document}